\documentclass[conference,10pt]{IEEEtran}
\usepackage{graphicx}
\usepackage{amsmath}
\usepackage{mathtools}
\usepackage{amssymb}
\usepackage{math_symbols}
\usepackage{calc}
\usepackage{booktabs,hhline} 
\usepackage{enumitem}
\usepackage{siunitx}
\usepackage[english]{babel}
\usepackage{times}
\usepackage{url}
\usepackage{tikz}
\usetikzlibrary{calc,arrows,positioning}

\usepackage{tabularx}

\newcommand\clearrow{\global\let\rowmac\relax}
\clearrow

\usepackage{cite} 

\def\tbfpi{\tilde{\bfpi}}

\def\bbfpi{\bar{\bfpi}}
\def\Sgaga{\bfSigma_{\bfgamma\bfgamma}}
\def\Sbgbg{\bfSigma_{\bbfg\bbfg}}
\def\Sgg{\bfSigma_{\bfg\bfg}}
\def\Sgu{\bfSigma_{\bfg\bfu}}
\def\Sug{\bfSigma_{\bfu\bfg}}
\def\Sff{\bfSigma_{\bff\bff}}
\def\Sfu{\bfSigma_{\bff\bfu}}
\def\Suf{\bfSigma_{\bfu\bff}}
\def\Suu{\bfSigma_{\bfu\bfu}}

\def\px{\mathsf{x}}
\def\py{\mathsf{y}}
\def\vx{\dot{\mathsf{x}}}
\def\vy{\dot{\mathsf{y}}}
\def\Sy{\bfS_{\bfy}}
\def\Su{\bfS_{\bfu}}
\def\Sv{\bfS_{\bfv}}
\def\oy{\bfo_{\bfy}}
\def\ou{\bfo_{\bfu}}
\def\atan{\mathrm{atan2}}

\begin{document}
\onecolumn
\vspace*{5cm}
This paper has been accepted for publication in 2024 27th International Conference on Information Fusion (FUSION)
Please cite the paper as:
O. Straka and J. Havlík, "Design of Unitless Normalized Measure of Nonlinearity for State Estimation," 2024 27th International Conference on Information Fusion (FUSION), Venice, Italy, 2024, pp. 1-7, doi: 10.23919/FUSION59988.2024.10706367. 
\clearpage
\twocolumn
\title{Design of Unitless Normalized Measure of Nonlinearity for State Estimation}

\author{\IEEEauthorblockN{Ond\v{r}ej Straka\IEEEauthorrefmark{1}, Jind\v{r}ich Havl\'{i}k\IEEEauthorrefmark{1},} \IEEEauthorblockA{\IEEEauthorrefmark{1}
    Dept.
  \ of Cybernetics, \& European Centre of Excellence NTIS \\Faculty of Applied Sciences, Univ.\ of West Bohemia, Czech Republic\\ E-mails: $\{$straka30,havlikj$\}$@kky.zcu.cz}
}%
\maketitle
\begin{abstract}
  The paper deals with measures of nonlinearity.
  In state estimation, they are utilized \emph{i)}~to select a suitable state estimation algorithm by assessing the nonlinearity of a system model, \emph{ii)}~to adapt the estimation algorithm structure or parameters, or \emph{iii)}~to indicate the possible effect of strong nonlinearity that leads to estimate credibility loss.
  This paper summarizes the state of the art of nonlinearity measures, focusing on the mean-square-error-based measure of nonlinearity.
  Its weak point is illustrated, and based on this, requirements for the new measure of nonlinearity are formulated.
  A new nonlinearity measure that is both unitless and normalized is designed.
  Its properties are demonstrated using numerical tracking experiments.
\end{abstract}

\section{Introduction}
Mathematical modeling of systems aims to describe the behavior of the system under study so that the behavior of individual components of the system can be better understood and predictions about future behavior can be made.
The simplest models exhibit proportionality, i.e., they are linear.
Linear problems can be solved efficiently and quickly.
Unfortunately, linear behavior is idealized, and models of real systems are not purely linear.
Moreover, the relationship between the variables often exhibits a character of uncertainty.
It is not difficult to recognize that the model or its component behavior is not linear, but the question is: By how much?
Is a linear approximation too simplifying?
Do linear methods provide a viable approach?
To answer these questions, one needs to be able to quantify the amount of deviation from linear behavior.
Given the importance of these issues, it is not surprising that methods capable of quantifying nonlinearity are found in every field employing a mathematical approach to modeling.
In the sequel, a method quantifying the nonlinearity will be called a measure of nonlinearity (MoN).

In 1960, Beale~\cite{Be:60} was the first one who theoretically dealt with MoNs and proposed empirical and theoretical MoNs based on the difference between the nonlinear function and the Taylor expansion for the estimation of static non-random parameters.
They were intended to indicate the accuracy of the approximate confidence intervals in regression analysis and to seek a linear approximation of the model.
Many reactions investigating the properties and behavior of these MoNs appeared, such as \cite{GuMe:65,Li:75}, which pointed out the inappropriate behavior of the MoNs for strongly nonlinear problems and proposed suitable modifications.
Based on principles from differential geometry, the so-called curvature\footnote{They assess deviation of the nonlinear function from a tangent (hyper)plane.
}
MoNs~\cite{BaWa:88} were proposed for estimating a static parameter.

In nonlinear deterministic dynamic system control, MoNs are used to investigate whether the linear control theory can be used.
For this purpose, MoNs based on functional norms of the difference between nonlinearity and the best linear approximation~\cite{DeWa:80,Ni:93}, MoNs using vector space norms of the output space along the system trajectory~\cite{HeMaAllgo:00}, or gap metric~\cite{Elsa:85} were designed.
MoNs are also applied in other fields such as time series~\cite{Schre:99, Ha:85}, clinical biochemistry~\cite{EmKr:93}, or econometrics~\cite{Ko:18}.

In the last few years, the MoNs have been the focus of intensive development in the state estimation of stochastic dynamical systems.
However, the task of state estimation is challenging for MoNs for several reasons: \emph{i)}~In addition to a static measurement model, the system models are also composed of an equation of state dynamics.
\emph{ii)}~The state is a random vector, and the nonlinearity has to be investigated in a specific region.
\emph{iii)}~Noise is an inherent part of stochastic models and has cumulative effects.

Differential MoNs were used to answer the question of whether the state should be modeled in a Cartesian or a polar coordinate system with radar~\cite{MaScaAru:05,MaSca:05} or video measurements\cite{MaSca:06}.
MoNs were used to decide whether a simple or more complex nonlinear filtering algorithm should be used~\cite{Pha:20}.
MoNs based on state density description by a mixture of Dirac functions were used to adapt nonlinear filter parameters to account for nonlinearity effects~\cite{StraDuSi_SYSID:12}.
Several MoNs have been designed benefiting from the fact that the linear transformations preserve the Gaussian distribution of the transformed variable.
These MoNs quantify the nonlinearity by the deviation of a variable density from the Gaussian distribution~\cite{DuStraGaFe:17}.
To this end, they use differential entropy, Kullback-Leibler divergence, various statistical tests, or higher-order moments.
There, the MoNs are utilized to adapt the filter structure parameters or to provide a filter with self-assessment of estimate consistency.
The paper~\cite{LiLi:15} proposed a unique MoN based on minimizing the mean square error (MSE) between a non-linear model influenced by a generally non-additive noise and its best linear approximation with an additive noise.
Compared with other MoNs, the MSE-based MoN offers many favorable properties, such as normalization, simultaneous assessment of state dynamics and measurement relations, and consideration of generally non-additive noises.

The MSE-based MoN was shown in~\cite{MaRi:17} to provide counter-intuitive results for different nonlinear estimation problems, as the MoN value was inconsistent with experts' expectations.
In addition, as will be shown in the motivation example, the MSE-based MoN behavior is highly sensitive to the choice of the variable units, which may be critical, especially for the MoN used in assessing the nonlinearity of different models.
Therefore, the paper aims to identify the shortcomings of the MoN, revise the requirements for MoNs in relation to the task model nonlinearity assessment, and design a new MoN that does not suffer from these deficiencies.

The rest of the paper is organized as follows.
Section~\ref{sec:problem_statement_and_mse_based_mon} defines the MoN and describes the MSE-based MoN introduced in~\cite{LiLi:15}.
The requirements for the new MoN are particularized in Section~\ref{sec:requirements_for_mon}, and the new unitless MoN is designed in Section~\ref{sec:unitless_mse_based_mon}.
The MoN is numerically illustrated in Section~\ref{sec:numerical_illustration}, and concluding remarks are given in Section~\ref{sec:conclusion}.
\section{Problem Statement and MSE-based MoN}\label{sec:problem_statement_and_mse_based_mon}
Consider a nonlinear vector transformation $\bfg:\real^{n_u}\times\real^{n_v}\mapsto \real^{n_y}$ from the space of random variable $\bfu\in\real^{n_u}$ and noise $\bfv\in\real^{n_v}$ to the space of random variable $\bfy\in\real^{n_y}$
\begin{align}
  \label{ali:g_general} \bfy=\bfg(\bfu,\bfv)
\end{align}
with known probability density functions (PDFs) $p(\bfu)$ and $p(\bfv)$.
The noise $\bfv$ is assumed without loss of generality zero-mean\footnote{In the case of a nonzero-mean noise, the mean can be embedded into the function $\bfg$.
} and independent of $\bfu$.
In discrete-time state space models 
\begin{subequations}\label{ali:state_model}
\begin{align}\label{ali:state_dynamics}
  \bfx_{k+1}&=\bff_k(\bfx_k,\bfw_k)\\\label{ali:measurement}
  \bfz_k&=\bfh_k(\bfx_k)+\bfv_k,
\end{align}
\end{subequations}
the transformation~\eqref{ali:g_general} may represent the state $\bfx_k$ dynamics~\eqref{ali:state_dynamics}, the relation~\eqref{ali:measurement} between the state and the measurement $\bfz_k$, or both if they are stacked.
Note that variables $\bfw_k$ and $\bfv_k$  represent the process and measurement noises, respectively.

The transformation~\eqref{ali:g_general} is the most general, with the noise~$\bfv$ acting non-additively.
Often, one is faced with a special case where the noise acts multiplicatively
\begin{align}
  \label{ali:g_multiplicative} \bfy=\bfg(\bfu,\bfv)=\bff(\bfu)+\bfpi(\bfu)\bfgamma(\bfv),
\end{align}
or additively
\begin{align}
  \label{ali:g_additive} \bfy=\bfg(\bfu,\bfv)=\bff(\bfu)+\bfv,
\end{align}
i.e., $n_v=n_y$ and $\bff:\real^{n_u}\mapsto \real^{n_y}$. The term $\bfgamma(\bfv)$ in~\eqref{ali:g_multiplicative}  is assumed zero mean without loss of generality, i.e., $\mean[\bfgamma(\bfv)]=\bfnul$. 
In the sequel, the following notation for covariance matrices is used: $\Sgg \coloneqq \var[\bfg(\bfu,\bfv)]$, $\Suu \coloneqq \var[\bfu]$, $\Sgu \coloneqq \cov[\bfg(\bfu,\bfv),\bfu]$, $\Sug=\Sgu\T$, $\Sff \coloneqq \var[\bff(\bfu)]$, $\Sfu \coloneqq \cov[\bff(\bfu),\bfu]$, $\Suf=\Sfu\T$, $\Sgaga \coloneqq \var[\bfgamma(\bfu)]$. 

The MoNs measure the deviation of~\eqref{ali:g_general} (or~\eqref{ali:g_multiplicative} or~\eqref{ali:g_additive}) from linear\footnote{Mathematically strictly speaking, the MoNs often measure deviation from affine behavior rather than from linear behavior.
} behavior.
The deviation can be assessed globally within the whole domain $\real^{n_u}\times\real^{n_v}$ of $\bfg$ or locally within a subset of the domain, where the subset is often determined by the PDFs $p(\bfu)$ and $p(\bfv)$.

Assume that the space of all mappings~\eqref{ali:g_general} is denoted by~$\calG$ and the space of the cumulative distribution functions (CDFs) corresponding to $ p(\bfu)$ and $p(\bfv)$ is denoted by~$\calD$.
Then, the MoN can be seen as a functional
\begin{align}
  \label{ali:M_general} M:\calG\times\calD\times\calD\mapsto \real_0^+,
\end{align}
which for a linear function $\bfg$ the functional returns zero.
Its value increases as $\bfg$ departs from linear behavior.
\subsection{MSE-based MoN}\label{sub:mse_based_mon}
The MoN based on the MSE was introduced in~\cite{Li:12} to quantify the nonlinearity of stochastic models in the form $\bfy=\bff(\bfu)$.
Later, it was generalized~\cite{LiLi:15} to mappings in the form~\eqref{ali:g_general}.
The MoN was based on the MSE of a difference between the nonlinear transformation and its best linear approximation.
\begin{align}\label{ali:MSEdefinition}
  M = \sqrt{\min_{\bfA,\bfb,\psi_{\bfv,\bfn}}\mean[\|\bfg(\bfu,\bfv)-\bfL(\bfu,\bfn)\|_2^2]}
\end{align}
where the expectation is done over all involved random variables, i.e., $\bfu$, $\bfv$, and $\bfn$ and the linear approximation $\bfL(\bfu,\bfn)$ is of the form
\begin{align}
  \bfL(\bfu,\bfn) = \bfA\bfu+\bfb+\bfn,
\end{align}
with $\bfA$ and $\bfb$ being matrix and vector of corresponding dimensions and $\bfn$ being a zero-mean noise.
The optimization in~\eqref{ali:MSEdefinition} is thus carried out over $\bfA$, $\bfb$, and the joint distribution $\psi_{\bfv,\bfn}$.

Also, a normalized version of~\eqref{ali:MSEdefinition} was introduced in~\cite{LiLi:15} in the form
\begin{align}\label{ali:MSENdefinition} 
  M^\text{norm}=\frac{M}{\sqrt{\tr(\Sgg)}}
\end{align}
such that $M^\text{norm}\in[0,1]$.
It was shown~\cite{Li:12} that for~\eqref{ali:g_general} the optimal values of $\bfA$ and $\bfb$ are
\begin{align}\label{ali:A}
  \bfA & =\bfSigma_{\bfg\bfu}\bfSigma_{\bfu\bfu}^{-1} \\\label{ali:b}
  \bfb & =\mean[\bfy]-\bfA\mean[\bfu].
\end{align}
For the general noise case~\eqref{ali:g_general}, the MoN cannot be simplified further and equals to
\begin{multline}
  M=  \sqrt{\tr(\bfSigma_{\bfg\bfg}-\bfSigma_{\bfg\bfu}\bfSigma_{\bfu\bfu}^{-1}\bfSigma_{\bfu\bfg})+} \\
      \overline{+\min_{\psi_{\bfv,\bfn}}(\mean\|\bfn\|_2^2-2\mean[\bfg(\bfu,\bfv)\T\bfn])}.
\end{multline}
For the multiplicative noise case~\eqref{ali:g_multiplicative}, the MoN equals to
\begin{align}
  M = \sqrt{\tr(\Sff-\Sfu\Suu^{-1}\Suf)+\tr(\mean[\tbfpi\T\tbfpi\Sgaga])}
\end{align}
and for the additive noise case~\eqref{ali:g_additive}, the MoN simplifies to
\begin{align}
  M = \sqrt{\tr(\Sff-\Sfu\Suu^{-1}\Suf)}.
\end{align}
\subsection{Motivation Example}\label{sub:motivation_example}
Consider the problem of transformation of a Gaussian distributed position from Cartesian to polar coordinates, i.e., $\bfu=[\px,\,\py]\T$ given by
\begin{align}
  \label{ali:u}
  \begin{bmatrix}
    \px \\
    \py
  \end{bmatrix}
   & \sim\calN\left(
     \begin{bmatrix}
       1 \\
       10
     \end{bmatrix}
     [\text{km}],\,\alpha\cdot
     \begin{bmatrix}
       1 & 0   \\
       0 & 100
     \end{bmatrix}
   [\text{km}^2]\right),
\end{align}
is transformed through either
\begin{align}
  \label{ali:motivate_deg} \bfy & =\bff(\bfu)=
  \begin{bmatrix}
    \sqrt{\px^2+\py^2} \\
    \atan(\py,\px)\tfrac{180}{\pi}
  \end{bmatrix}
  \begin{matrix} [\text{km}] \\ [\,^\circ]
  \end{matrix}
\end{align}
or
\begin{align}
  \label{ali:motivate_rad} \bfy & =\bff(\bfu)=
  \begin{bmatrix}
    \sqrt{\px^2+\py^2} \\
    \atan(\py,\px)
  \end{bmatrix}
  \begin{matrix} [\text{km}] \\ [\text{rad}]
  \end{matrix}
\end{align}
The transformations~\eqref{ali:motivate_deg} and~\eqref{ali:motivate_rad} differ only in the units for the bearing.

The values of $\alpha$ in~\eqref{ali:u} will be selected in the interval $\alpha\in[10^{-4},10^2]$.
Changing the values of $\alpha$ facilitates the analysis of the MoN since the parameter affects the region over which the nonlinearity of $\bff$ is studied.
The region is small for small values of $\alpha$, and the function will be close to linear in such a region.
Thus, changing $\alpha$ indirectly changes the nonlinearity.
Changing the nonlinearity of $\bff$ directly is difficult.

\begin{figure}
  \centering
  \includegraphics[width=\linewidth]{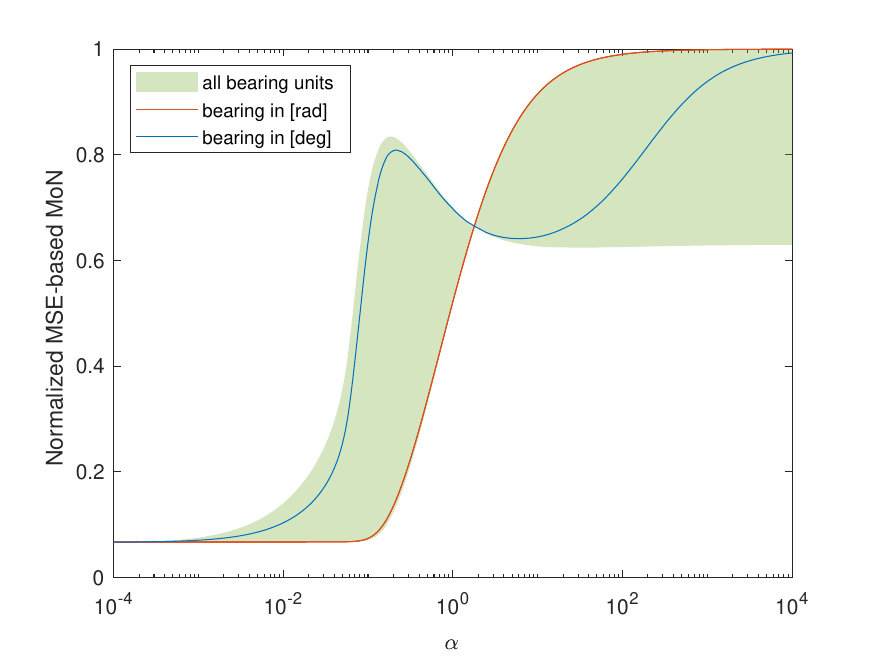}
  \caption{Effect of various bearing units on the MSE-based MoN.}
  \label{fig:illustration}
\end{figure}

The normalized MSE-based MoN~\cite{LiLi:15}
\begin{align}
  M^\text{norm} = \frac{\sqrt{\tr(\Sff-\Sfu\Suu^{-1}\Suf)}}{\sqrt{\tr(\Sff)}}
\end{align}
was calculated for both~\eqref{ali:motivate_deg} and~\eqref{ali:motivate_rad} and is shown in Figure~\ref{fig:illustration}.
The lines for~\eqref{ali:motivate_rad} and~\eqref{ali:motivate_deg} indicate huge difference between the MoN values for some $\alpha$ (for $\alpha=0.17$ the MoN for bearing in radians is $ M^\text{norm}=10\%$ while it is $M^\text{norm}=80\%$ for bearing in degrees).
In addition, Figure~\ref{fig:illustration} also depicts the area of MoN values for arbitrary linear change of bearing units, (i.e., from units equivalent to \SI{10}{\giga\radian} to units equivalent to \SI{0.1}{\nano\radian}).
To summarize, the illustration example indicates a significant dependence of the MSE-based MoN on the units of $\bfy$, which motivates the design of a unitless MoN.
\section{Requirements for MoN}\label{sec:requirements_for_mon}
Motivated by the illustration and the discussion above, the designed MoN should possess the following properties:
\begin{description}
  \item[P1:] The MoN should be unitless, i.e., its value should be invariant to a change of units of $\bfy$ and $\bfu$.
  \item[P2:] The MoN should be normalized, i.e., provide values in $[0,1]$.
\end{description}
The property P1 can expressed through the invariance of MoN to a change of units.
The change of units of quantities $\bfu$, $\bfy$ , and $\bfv$  is assumed to be an affine transformation, i.e.,
\begin{subequations}
  \label{ali:transformations}
  \begin{align}
    \bfu & =\Su\bbfu+\ou  \\
    \bfy & =\Sy\bbfy+\oy, \\
    \bfv & =\Sv\bbfv,
  \end{align}
\end{subequations}
where $\bbfu$, $\bbfy$, and $\bbfv$ are the quantities $\bfu$, $\bfy$, and $\bfv$, respectively, expressed in some base units, $\Su\in\real^{n_u\times n_u}$, $\Sy\in\real^{n_y\times n_y}$, and $\Sv\in\real^{n_v\times n_v}$ are positive definite \textit{scale} matrices and $\ou\in\real^{n_u}$ and $\oy\in\real^{n_y}$ are offset vectors\footnote{Note that the noise $\bbfv$ still has zero mean.
  No offset is needed to transform $\bbfv$ to $\bfv$.
}.
To regularize the problem, the base units will be chosen such that their elements have variances equal to one, i.e., $\diag(\Sbgbg)=\bfone_{n_y\times 1}$, $\diag(\bfSigma_{\bbfu\bbfu})=\bfone_{n_u\times 1}$, and $\diag(\bfSigma_{\bbfv\bbfv})=\bfone_{n_v\times 1}$, where $\bfone_{a\times b}$ represents an $ a\times b$  matrix of ones.
Analogously to~\eqref{ali:g_general}, the base units are related through a mapping $\bbfg$ as 
\begin{align}
  \bbfy = \bbfg(\bbfu,\bbfv)
\end{align}
and thus
\begin{align}
  \bfy=\bfg(\bfu,\bfv)=\Sy\bbfg\left(\Su^{-1}(\bfu-\ou),\Sv^{-1}\bfv\right)+\oy.
\end{align}
The property P1 can then be written as
\begin{align}
  M(\bfg,\psi_\bfu,\psi_\bfv)=M(\bbfg,\psi_{\bbfu},\psi_{\bbfv}),
\end{align}
where $\psi_\bfu$, $\psi_\bfv$, $\psi_{\bbfu}$, and $\psi_{\bbfv}$ are distributions of $\bfu$, $\bfv$, $\bbfu$, and $\bbfv$, respectively.
Note that the MoN~\eqref{ali:MSEdefinition} was shown to be independent~\cite{LiLi:15} of the affine transformation of $\bfu$ and $\bfv$.

The property P2 can be expressed simply as
\begin{align}
  \label{ali:M_normalized} M:\calG\times\calD\times\calD\mapsto [0,1].
\end{align}
\section{Unitless MSE-based MoN}\label{sec:unitless_mse_based_mon}
To achieve the unitless property of the MSE-based MoN~\eqref{ali:MSENdefinition}, the $L_2$ norm $\|\cdot\|_2$ is replaced by a weighted norm $\|\cdot\|_\bfW=(\cdot)\bfW(\cdot)\T$, where $\bfW\in\real^{n_y\times n_y}$ is a positive definite matrix.
\begin{align}
  \label{ali:UMSEdefinition}
  M = \sqrt{\min_{\bfA,\bfb,\psi_{\bfv,\bfn}}\mean[\|\bfg(\bfu,\bfv)-\bfL(\bfu,\bfn)\|_\bfW]}
\end{align}
Note that using the weighted norm was mentioned in~\cite{Li:12} as an option without further details.

\subsection{Minimization}\label{sub:minimization}
The assumptions used for the minimization in~\eqref{ali:UMSEdefinition} are:
\begin{description}
  \item[A1:] Variable $\bfu$ is independent of $\bfv$ and $\bfn$, i.e. $\psi_{\bfu,\bfv,\bfn}=\psi_{\bfu}\psi_{\bfv,\bfn}$.
  \item[A2:] The minimization in~\eqref{ali:UMSEdefinition} is done  w.r.t.\ $\bfA$, $\bfb$  and all joint distributions $\psi_{\bfn,\bfv}$  such that the marginal  distribution $\psi_{\bfv}$ is fixed and $\bfn$  has zero mean.
\end{description}
From \textbf{A2}, it follows that the minimization is done w.r.t.
\ conditional distribution $\psi_{\bfn|\bfv}$ as $\psi_{\bfn,\bfv}=\psi_{\bfn|\bfv}\psi_{\bfv}$ and $\psi_\bfv$ is fixed.
The minimization can thus be rewritten as
\begin{align}
  \label{ali:minimization} & \min_{\bfA,\bfb,\psi_{\bfn|\bfv}}\mean[\|\bfg(\bfu,\bfv)-(\bfA\bfu+\bfb+\bfn)\|_\bfW]                                \\
  \nonumber =              & \min_{\bfA,\bfb,\psi_{\bfn|\bfv}}\Big\{\mean[\|\bfg(\bfu,\bfv)-(\bfA\bfu+\bfb)\|_\bfW]+\mean[\|\bfn\|_\bfW]          \\
  \nonumber                & \qquad -2\mean[(\bfg(\bfu,\bfv)-\bfA\bfu-\bfb)\bfW\bfn\T]\Big\}                                                             \\
  \nonumber =              & \min_{\bfA,\bfb}\mean[\|\bfg(\bfu,\bfv)-(\bfA\bfu+\bfb)\|_\bfW]                                                      \\
  \nonumber                & \qquad +\min_{\psi_{\bfn|\bfv}}\Big\{\mean[\|\bfn\|_\bfW]-2\mean[(\bfg(\bfu,\bfv)\bfW\bfn\T]\Big\}                          \\
  \nonumber =              & \underbrace{\min_{\bfA,\bfb}\mean[\|\bfg(\bfu,\bfv)-(\bfA\bfu+\bfb)\|_\bfW]}_{J_{\bfA,\bfb}}                         \\
  \nonumber                & \qquad +\underbrace{\min_{\psi_{\bfn|\bfv}}\mean[\|\bfg(\bfu,\bfv)-\bfn\|_\bfW]}_{J_{\bfn}}-\mean[\|\bfg(\bfu,\bfv)\|_\bfW].
\end{align}
Using~\eqref{ali:minimization}, the MoN~\eqref{ali:UMSEdefinition} equals to
\begin{align}
  \label{ali:UMSEvalue} M=\sqrt{J_{\bfA,\bfb}+J_{\bfn}-\mean\left[\|\bfg(\bfu,\bfv)\|_{\bfW}\right]},
\end{align}
which means that the minimization in~\eqref{ali:UMSEdefinition} was split into the deterministic part $J_{\bfA,\bfb}$ and the stochastic part $J_{\bfn}$.
The deterministic minimization can easily be solved~\cite{Li:12} and leads to
\begin{align}
  \label{ali:Dmimimization} J_{\bfA,\bfb}=\tr\left(\bfW(\Sgg-\Sgu\Suu^{-1}\Sug)\right),
\end{align}
where $\bfA$  and $\bfb$  are given by~\eqref{ali:A} and \eqref{ali:b}, respectively.

On the other hand, the minimization $J_{\bfn}$ cannot be done analytically for the general transformation~\eqref{ali:g_general}.
For the multiplicative form~\eqref{ali:g_multiplicative}, the minimization leads to
\begin{align}
  \label{ali:Sminimization} J_{\bfn}=\mean[\|\bfg(\bfu,\bfv)\|_{\bfW}]-\tr(\bfW\bbfpi\Sgaga\bbfpi\T),
\end{align}
where $\bbfpi=\mean[\bfpi(\bfu)]$.
For details, see Appendix~\ref{sec:app_jn}.
Thus, for the multiplicative noise case~\eqref{ali:g_multiplicative}, the MoN is
\begin{align}
  \label{ali:UMSEmultiplicative} M & = \sqrt{\tr\left(\bfW(\Sgg-\Sgu\Suu^{-1}\Sug-\bbfpi\Sgaga\bbfpi\T)\right)}                    \\
  \nonumber                        & = \sqrt{\tr\left(\bfW(\Sff-\Sfu\Suu^{-1}\Suf+\mean[\widetilde{\bfpi}(\bfu)\Sgaga\widetilde{\bfpi}(\bfu)\T)\right)},
\end{align}
where the relations $\Sgu=\Sfu$, $\Sgg=\Sff+\mean[\bfpi(\bfu)\Sgaga\bfpi(\bfu)\T]$, and $\widetilde{\bfpi}(\bfu)=\bfpi(\bfu)- \bbfpi$ were used.
For the additive noise case~\eqref{ali:g_additive}, the MoN is
\begin{align}
  \label{ali:UMSEadditive} M & =\sqrt{\tr\left(\bfW(\Sff-\Sfu\Suu^{-1}\Suf\right)}.
\end{align}
\subsection{Normalization of MoN}\label{sub:normalization_of_mon} The normalization of the MoN~\eqref{ali:UMSEvalue} is based on finding an upper bound of $J_{\bfA,\bfb}$ and $ J_{\bfn}$.
To find the upper bound of $J_{\bfA,\bfb}$, no linearity in $\bfg$ is assumed, i.e., $\bfA=\bfnul$ and the optimum value of the shift~\eqref{ali:b} is used to obtain a tight upper bound, i.e.,
\begin{multline}
  \label{ali:Jab_bound} J_{\bfA,\bfb}= \min_{\bfA,\bfb}\mean[\|\bfg(\bfu,\bfv)-(\bfA\bfu+\bfb)\|_\bfW]\\ \leq\mean\left[\|\bfg(\bfu,\bfv)-\mean[\bfg(\bfu,\bfv)]\|_\bfW\right]=\tr(\bfW\Sgg).
\end{multline}
The upper bound of the stochastic part $J_\bfn$ can be found by rewriting the norm in $J_\bfn$ as
\begin{multline}
  \label{ali:norm} \mean[\|\bfg(\bfu,\bfv)-\bfn\|_\bfW]=\mean[\|\bfg(\bfu,\bfv)\|_\bfW+\|\bfn\|_\bfW\\-2\bfg(\bfu,\bfv)\bfW\bfn\T].
\end{multline}
When $\bfv$ and $\bfn$ are independent, the expectation of the last term is zero as $\mean[\bfn]=\bfnul$.
Also, a distribution $\psi_\bfn$ must exist such that $\mean[\|\bfn\|_\bfW]=0$.
Then, the upper bound of $ J_\bfn$ is
\begin{align}
  \label{ali:Jn_bound} J_\bfn\leq\mean[\|\bfg(\bfu,\bfv)\|_\bfW].
\end{align}
Substituting \eqref{ali:Jab_bound} and \eqref{ali:Jn_bound} to \eqref{ali:UMSEvalue} then results in
\begin{align}
  \label{ali:M_bound} M=\sqrt{J_{\bfA,\bfb}+J_{\bfn}-\mean[\|\bfg(\bfu,\bfv)\|_\bfW]}\leq\sqrt{\tr(\bfW\Sgg)}.
\end{align}
Note that the upper bound~\eqref{ali:M_bound} is achieved for zero correlation $\Sfu=\bfnul$ in the additive case~\eqref{ali:g_additive} and for the zero correlation and $\mean[\widetilde{\bfpi}(\bfu)\Sgaga\widetilde{\bfpi}(\bfu)\T]=\bfnul$ in the multiplicative case~\eqref{ali:g_multiplicative}.

Specific choices of the weighting matrix will be discussed in the following section.
\subsection{Weighting Matrix Choice}\label{sub:weighting_matrix_choice}
The unitless property of the MoN~\eqref{ali:UMSEdefinition} depends on the weight matrix $\bfW$, i.e., the following equality must hold
\begin{align}
  \label{ali:mon_equality} M(\bfg,\psi_\bfu,\psi_\bfv;\bfW)= M(\bbfg,\psi_{\bbfu},\psi_{\bbfv};\bbfW),
\end{align}
where the notation explicitly expresses the MoN as a functional parametrized by the corresponding weight matrix.
Using the relations~\eqref{ali:transformations} it follows that
\begin{subequations}
  \label{ali:transformations_ii}
  \begin{align}
    \bbfg(\bbfu,\bbfv)  & =\Sy^{-1}\bfg(\Su\bbfu+\ou,\Sv\bbfv)-\Sy^{-1}\oy \\
    \psi_{\bbfu}(\bbfu) & =\psi_{\bfu}(\Su\bbfu + \ou)                     \\
    \psi_{\bbfv}(\bbfv) & =\psi_{\bfv}(\Sv\bbfv)
  \end{align}
\end{subequations}
Using the notation $\widetilde{\bfg}(\bfu,\bfv)=\Sy^{-1}\bfg(\bfu,\bfv)-\Sy^{-1}\oy$, the MoN for $\bbfg$ can be rewritten as
\begin{align}
  \label{ali:mon_equality_ii} M(\bbfg,\psi_{\bbfu},\psi_{\bbfv};\bbfW)=M(\widetilde{\bfg},\psi_{\bfu},\psi_{\bfv};\bbfW)
\end{align}
Using simple manipulations, it can be shown that
\begin{align}
  \label{ali:mon_equality_iii} M(\widetilde{\bfg},\psi_{\bfu},\psi_{\bfv};\bbfW)=M(\bfg,\psi_{\bfu},\psi_{\bfv};\Sy^{-1}\bbfW\Sy^{-1}),
\end{align}
Combining~\eqref{ali:mon_equality}, \eqref{ali:mon_equality_ii}, and \eqref{ali:mon_equality_iii} it can be concluded that to achieve the property P1, the weights must be related as
\begin{align}
  \label{ali:weight_relation} \bfW=\Sy^{-1}\bbfW\Sy^{-1}.
\end{align}

The equality between the weight matrices~\eqref{ali:weight_relation} and the normalization condition $\tr(\bfW\Sgg)=\tr(\bbfW\Sbgbg)=1$ following from~\eqref{ali:M_bound} can be used to obtain matrix $\bfW$ leading to a unitless and normalized MoN.

Recalling $\diag(\Sbgbg)=\bfI$, simple choices of weight matrices $\bbfW$ for the base units following directly from the normalization condition $\tr(\bbfW\Sbgbg)=1$ are
\begin{align}
  \label{ali:bw_diag_choice} \bbfW^\text{diag} & =\frac{1}{n_y}\bfI         \\
  \label{ali:bw_full_choice} \bbfW^\text{full} & =\frac{1}{n_y}\Sbgbg^{-1}.
\end{align}

The complete set of weight matrices $\bbfW$ satisfying $\tr(\bbfW\Sbgbg)=1$ is given\footnote{The notation $[\cdot]_{ii}$ stands for the element at the $i$-th row and $i$-th column.
} by
\begin{align}
  \label{ali:bweight_set}
  \bbfW\in\{\bfV\bfY\bfV\T;\bfY\in\bfS_{++}^{n_y},[\bfY]_{ii}=\tfrac{\alpha_i}{\lambda_i}, i=1\dots n_y\},
\end{align}
where $\Sbgbg=\bfV\bfLambda\bfV\T$, $\bfV$ is the matrix of eigenvectors of $\Sbgbg$, $\bfLambda$  is the matrix of eigenvalues, $\lambda_i=[\bfLambda]_{ii}$, $\alpha_i>0$  are parameters satisfying $\sum_{i=1}^{n_y}\alpha_i=1$  and $\bfS_{++}^{n_y}=\{\bfX\in\bfS_{++}^{n_y};\bfX\succ\bfnul\}$ is a set of $n_y$-dimensional positive definite matrices.
For the explanation, see Appendix~\ref{sec:app_weights}.

The weights~\eqref{ali:bweight_set}, for which~\eqref{ali:bw_diag_choice} an \eqref{ali:bw_full_choice} are special cases, ensure normalization of the MoN and achieve the unitless property.
Note that both $\|\hbfg(\bfu,\bfv)-\bfL(\bfu,\bfn)\|_{\tfrac{1}{n_y}\bfI}$ and $\|\hbfg(\bfu,\bfv)-\bfL(\bfu,\bfn)\|_{\tfrac{1}{n_y}\Sbgbg^{-1}}$ are unitless because $\bfI=\diag(\diag(\Sbgbg))^{-1}$ is an inverse of a diagonal matrix with the diagonal\footnote{Here, the MATLAB notation is used in the expression.
} given by $\diag(\Sbgbg)$.

The weight matrices $\bfW$ for the original units are then obtained by~\eqref{ali:weight_relation} with~\eqref{ali:bw_diag_choice} or \eqref{ali:bw_full_choice} in the form
\begin{align}
  \label{ali:w_diag_choice} \bfW^\text{diag} & =\frac{1}{n_y}\Sy^{-1}\Sy^{-1}=\frac{1}{n_y}\diag(\diag(\Sgg))^{-1} \\
  \label{ali:w_full_choice} \bfW^\text{full} & =\frac{1}{n_y}\Sy^{-1}\Sbgbg^{-1}\Sy^{-1}=\frac{1}{n_y}\Sgg^{-1}.
\end{align}

The MoN~\eqref{ali:UMSEdefinition} computed by~\eqref{ali:UMSEmultiplicative} for the multiplicative noise case and by~\eqref{ali:UMSEadditive} for the additive noise case with weight matrix $\bfW=\Sy^{-1}\bbfW\Sy^{-1}$, where $\bbfW$ is given by~\eqref{ali:bweight_set} constitutes the desired unitless normalized MoN.
The individual weight matrices, such as~\eqref{ali:w_diag_choice} or~\eqref{ali:w_full_choice}, differ in the way they treat the correlations among individual elements of $\bfy$.
For example, the MoN with the choice~\eqref{ali:w_diag_choice} ignores the off-diagonal elements of $(\Sff-\Sfu\Suu^{-1}\Suf)$ while the MoN with the choice~\eqref{ali:w_full_choice} weighs them through $\Sgg^{-1}$. 


\textit{Remark 1:} When considering the transformation~\eqref{ali:g_general} without the noise, i.e., $\bfy=\bfg(\bfu)$, the MoN~\eqref{ali:UMSEdefinition} with $\bfW^\text{full}$  corresponds to square root of the multivariate generalization of coefficient of determination~\cite{ReChri:12} which has many convenient features.

\textit{Remark 2:}
For the weight matrix $\bfW=\bfI$ the original MSE-based norm~\eqref{ali:MSEdefinition} is obtained and the weight matrix $\bfW=\tfrac{1}{\sqrt{\tr(\Sgg)}}\bfI$ leads to the normalized MoN~\eqref{ali:MSENdefinition}.
However, none of the choices satisfy the property P1, i.e., they are not unitless.

\section{Numerical Illustration}\label{sec:numerical_illustration}
To illustrate the proposed unitless MoN, measurement functions of three tracking problems are considered: bearing-only tracking (BOT)~\cite{AidaHa:83,AruRi:00}, ground moving target indicator (GMTI) filtering~\cite{KiBaPaKa:00,WaBlaLiJoHaYiBeBre:16}, and tracking with range and direction cosine (RDCOS) measurements~\cite{Li:12}.
Analysis of MoNs for these problems was provided in~\cite{MaRi:17} with a focus on the normalized MSE-based MoN~\eqref{ali:MSENdefinition}.

\begin{figure*}
  \centering
  \includegraphics[width=\textwidth]{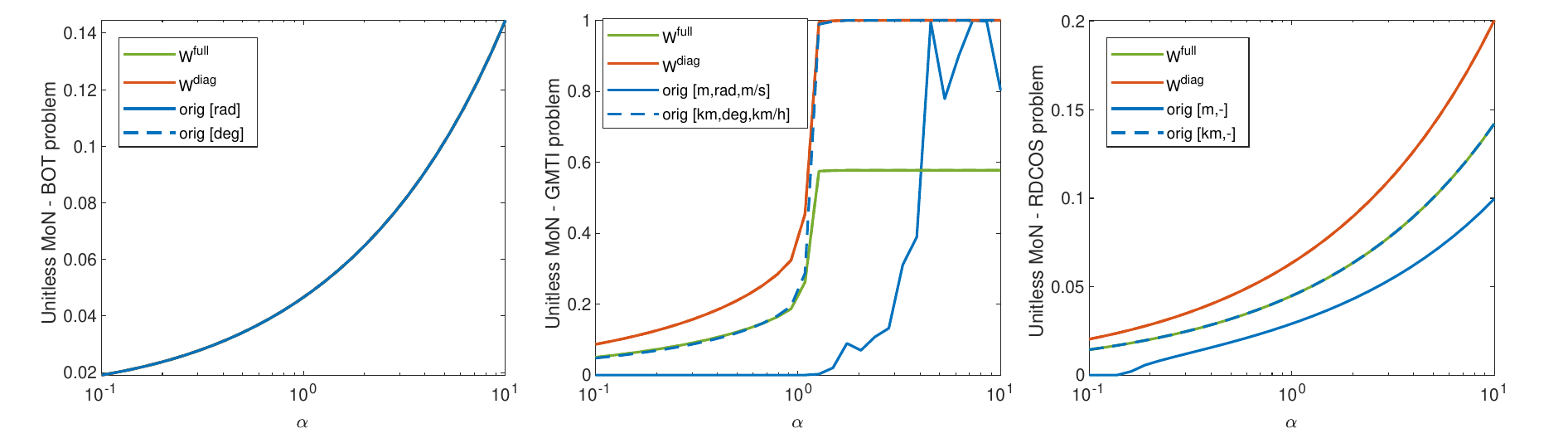}
  \caption{Comparison of MoNs for BOT, GMTI, and RDCOS problems: unitless MoN with $\bfW^\text{full}$ (green), unitless MoN with $\bfW^\text{diag}$ (red), original MSE-based MoN (blue and dashed blue).}
  \label{fig:example}
\end{figure*}

Assuming the Cartesian state of the target $\bfx=[\px,\,\py,\,\vx,\,\vy]\T$ with units $[\text{m},\,\text{m},\,\text{m/s},\,\text{m/s}]$ distributed as
\begin{align}
  \label{ali:example_x} \bfx\sim\calN\left\{[500,\,500,\,5,\,8.7]\T,\alpha\diag([10^3,\,10^3,\,1,\,1])\right\}
\end{align}
For the BOT, GMTI, and RDCOS the sensors provide measurements in the form
\begin{align}
  \label{ali:BOT} \bfz^\text{BOT}=\bfh^\text{BOT}(\bfx)=\atan(\px,\py),
\end{align}
\begin{subequations}
  \label{ali:GMTI}
  \begin{align}
    \label{ali:GMTIz} \bfz^\text{GMTI} & =[h_r(\bfx,\bfs),\, h_b(\bfx,\bfs),\, h_{v_r}(\bfx,\bfs)]\T,
  \end{align}
  \begin{align}
    \label{ali:GMTIr} h_r(\bfx,\bfs) & =\sqrt{(\px-s_x)^2+(\py-s_y)^2+(s_z)^2},
  \end{align}
  \begin{align}
    \label{ali:GMTIb} h_b(\bfx,\bfs) & =\atan(\px-s_x,\py-s_y),
  \end{align}
  \begin{align}
    \label{ali:GMTIvr} h_{v_r}(\bfx,\bfs) & =\frac{(\px-s_x)\vx+(\py-s_y)\vy}{h_r(\bfx,\bfs)},
  \end{align}
\end{subequations}
\begin{align}
  \label{ali:RDCOS} \bfz^\text{RDCOS}=\bfh^\text{RDCOS}(\bfx)=
  \begin{bmatrix}
    \sqrt{\px^2+\py^2} \\
    \px/\sqrt{\px^2+\py^2}
  \end{bmatrix}
  ,
\end{align}
where $\bfs=[s_x,\,s_y,\,s_z]\T=[10^3,\, 10^3,\, 10^3]\T$ is GMTI sensor position.
The value of the scaling factor $\alpha$ was set to $\alpha\in[10^{-1},10^1]$.
Three MoNs were calculated for the considered problems:
\begin{itemize}
  \item unitless MoN~\eqref{ali:UMSEadditive} with the full weight matrix~\eqref{ali:w_full_choice},
  \item unitless MoN~\eqref{ali:UMSEadditive} with the diagonal weight matrix~\eqref{ali:w_diag_choice},
  \item the original MoN~\eqref{ali:MSENdefinition} (denoted as orig).
\end{itemize}
The covariance matrices required to calculate the MoNs were obtained by the Monte Carlo method with $10^7$ samples.
The MoNs are depicted in Figure~\ref{fig:example}.
The figure shows that for the BOT problem, all MoN values coincide.
This is caused by the fact that $\bfz^\text{BOT}\in\real$ is scalar.
For such case $\bfW^\text{full}=\bfW^\text{diag}$.
Also, the value of the original MoN is equal to the proposed unitless MoN since for the scalar case dividing by the trace of a scalar in~\eqref{ali:MSENdefinition} corresponds to multiplying by $\bfW^\text{full}=\frac{1}{1}\Sff^{-1}$ in~\eqref{ali:UMSEadditive}.

For the GMTI and RDCOS problems, the values of the original MoN and the proposed unitless MoN differ.
To analyze sensitivity of the original MoN on the units, $\bfz^\text{GMTI}$ was considered either in $[\text{m},\,\text{rad},\, \text{m/s}]\T$ or $[\text{km},\,\text{deg},\, \text{km/h}]\T$ and $\bfz^\text{RDCOS}$ was considered in $[\text{m},-]$ or $[\text{km},-]$.
The figure shows that the original MoN may vary for the GMTI between $M=0$ for one set of units and $M=1$ for the other set of units (c.f.\
values of $\alpha$ close to one).
For the RDCOS problem, the differences are not so prominent.

The figure also shows that the unitless MoN's value depends on the weight matrix's choice $\bfW$.
The weight matrix $\bfW^\text{diag}$ leads to higher values of MoN in the considered example because the difference $(\Sff-\Sfu\Suu^{-1}\Suf)$ is positively correlated.
Generally, the unitless MoN with $\bfW^\text{diag}$ can be either lower or higher than the unitless MoN with $\bfW^\text{full}$.

Last, it shall be noticed that the MoN values for the BOT problem (interval $[2\times 10^{-1},\, 15\times 10^{-1}]$) are slightly smaller than the MoN values for the RDCOS problem (interval $[0,\,0.2]])$.
The MoN values for the GMTI are significantly higher (interval $[0,\,1]$), which indicates that compared to the BOT and RDCOS problems, the measurements of the GMTI exhibit strongly nonlinear behavior.
This aligns with the findings in~\cite{MaScaAru:05}.
\section{Concluding Remarks}\label{sec:conclusion}
The paper focused on measures of nonlinearity for the state estimation problem.
The measures of nonlinearity assess how far a transformation of random variables such as state and noise is from a linear behavior.
The mean square error-based measure of nonlinearity was analyzed, and it was shown to suffer from sensitivity in the choice of the units.
A new unitless nonlinearity measure was designed using a weight matrix to achieve the unitless property and normalization.
A general set of weight matrices was specified, and two distinctive weight matrix choices were presented.
The behavior of the new measure of nonlinearity was illustrated using the bearing-only tracking problem, ground target moving indicator filtering, and tracking with range and direction cosine.
\section{Acknowledgement}
This research was co-funded by the European Union under the project Robotics and Advanced Industrial Production (reg.\ no.\ CZ.02.01.01/00/22\_008/0004590).
\addtolength{\textheight}{-8cm}

\appendices
\section{Minimization of $J_\bfn$ }\label{sec:app_jn}
 To calculate 
 $J_{\bfn}=\min_{\psi_{\bfn|\bfv}}\mean[\|\bfg(\bfu,\bfv)-\bfn\|_\bfW]$
for the multiplicative case $\bfg(\bfu,\bfv)=\bff(\bfu)+\bfpi(\bfu)\bfgamma(\bfv)$,
it should be reminded that $\mean[\bfn]=\bfnul$, $\mean[\bfgamma(\bfv)]=\bfnul$, and $\bfu$ and $\bfv$  are assumed independent.
Then
\begin{multline}\label{ali:minJn_i}
  J_{\bfn}=\min_{\psi_{\bfn|\bfv}}\mean[\|\bff(\bfu)+\bfpi(\bfu)\bfgamma(\bfv)-\bfn\|_\bfW]\\
          =\mean\|\bff(\bfu)\|_{\bfW}+\min_{\psi_{\bfn|\bfv}}\mean[\|\bfpi(\bfu)\bfgamma(\bfv)-\bfn\|_{\bfW}].
\end{multline}
Now, by denoting $\bbfpi\coloneqq\mean[\bfpi(\bfu)]$, the term $\bfpi(\bfu)$  can be written as $\bfpi(\bfu)=\bbfpi+\widetilde{\bfpi}(\bfu)$. 
It follows that
\begin{multline}\label{ali:minJn_ii}
  \min_{\psi_{\bfn|\bfv}}\mean[\|\bfpi(\bfu)\bfgamma(\bfv)-\bfn\|_{\bfW}]\\=\min_{\psi_{\bfn|\bfv}}\left\{\mean[\|\widetilde{\bfpi}(\bfu)\bfgamma(\bfv)\|_{\bfW}] +\mean[\|\bbfpi\bfgamma(\bfv)-\bfn\|_{\bfW}]\right\}\\
 =\mean[\|\widetilde{\bfpi}(\bfu)\bfgamma(\bfv)\|_{\bfW}]+\min_{\psi_{\bfn|\bfv}}\mean[\|\bbfpi\bfgamma(\bfv)-\bfn\|_{\bfW}]\\
 =\tr\bfW\mean[\widetilde{\bfpi}(\bfu)\Sgaga\widetilde{\bfpi}\T],
\end{multline}
where $\min_{\psi_{\bfn|\bfv}}\mean[\|\bbfpi\bfgamma(\bfv)-\bfn\|_{\bfW}]=\bfnul$ for a distribution $\psi_{\bfn|\bfv}$ of $\bfn$ such that $\bfn=\bbfpi\bfgamma(\bfv)$  with probability one.

 By expressing $\mean[\|\bfg(\bfu,\bfv)\|_{\bfW}$  as
 \begin{multline}\label{ali:minJn_iii}
   \mean[\|\bfg(\bfu,\bfv)\|_{\bfW}]=\mean[\|\bff(\bfu)+(\bbfpi+\widetilde{\bfpi}(\bfu))\bfgamma(\bfv)\|_{\bfW}]\\
   =\mean[\|\bff(\bfu)\|_{\bfW}]+\tr\bfW\mean[\widetilde{\bfpi}(\bfu)\Sgaga\widetilde{\bfpi}(\bfu)\T] + \tr\bfW\bbfpi\Sgaga\bbfpi\T,
 \end{multline}
 the stochastic part $ J_\bfn$  is
\begin{multline}\label{ali:minJn_iv}
  J_{\bfn}=\min_{\psi_{\bfn|\bfv}}\mean[\|\bfg(\bfu,\bfv)-\bfn\|_\bfW]\\
  =\mean[\|\bfg(\bfu,\bfv)\|_{\bfW}]-\tr\bfW\bbfpi\Sgaga\bbfpi\T.
\end{multline}
\section{Set of weights}\label{sec:app_weights}
To find the set of matrices $\bbfW$ such that $\tr(\bbfW\Sbgbg)=1$, where $\bbfW\succ\bfnul$ and $\Sbgbg\succ\bfnul$, the eigen decomposition of $\Sbgbg=\bfV\bfLambda\bfV\T$ is computed first.
Then,
\begin{align*}
\tr(\bbfW\Sbgbg)=\tr(\bbfW\bfV\bfLambda\bfV\T)=\tr(\bfV\T\bbfW\bfV\bfLambda)=\tr(\bfY\bfLambda),
\end{align*}
where $\bfY=\bfV\T\bbfW\bfV$.
Since $\bfLambda=\diag([\lambda_1,\dots\lambda_{n_y}])$ is a diagonal matrix, the trace will equal to one, $\tr(\bfY\bfLambda)=1$, if for the diagonal elements of $[\bfY]_{ii}$ the following equality holds:
\begin{align*} [\bfY]_{ii}=\frac{\alpha_i}{\lambda_i},\, i=1,\dots,n_y,
\end{align*}
where $\sum_{i=1}^{n_y}=1,\, \alpha_i\geq 0$.
The off-diagonal elements $\bfY$ can be arbitrarily chosen such that $\bfY\succ\bfnul$.
The weight matrix $\bbfW$ then is $\bbfW=\bfV\bfY\bfV\T$.

\end{document}